\documentclass[pdflatex,sn-mathphys-num]{sn-jnl}% Math and Physical Sciences Numbered Reference Style

\usepackage{graphicx}%
\usepackage{multirow}%
\usepackage{amsmath,amssymb,amsfonts}%
\usepackage{amsthm}%
\usepackage{mathrsfs}%
\usepackage[title]{appendix}%
\usepackage{xcolor}%
\usepackage{textcomp}%
\usepackage{manyfoot}%
\usepackage{booktabs}%
\newcommand\blfootnote[1]{%
    \begingroup
    \renewcommand\thefootnote{}%
    \footnote{#1}%
    \addtocounter{footnote}{-1}%
    \endgroup
}
\usepackage{url}
\usepackage{dsfont}
\DeclareMathOperator*{\argmin}{\arg\!\min}
\usepackage[ruled,norelsize]{algorithm2e}
\usepackage{algpseudocode}%
\usepackage{listings}%
\usepackage{array}
\usepackage{mathtools}
\usepackage{threeparttable}
\usepackage{cellspace} % For adding vertical spacing in cells
\usepackage{makecell}
\usepackage{hhline}

\usepackage{rotating}
\usepackage{booktabs}

\AtBeginDocument{%
}

\newcommand{\codeurl}[1]{%
  \href{#1}{{\fontfamily{zi4}\selectfont #1}}%
}

\makeatletter
\let\cline\org@cline
\makeatother

\theoremstyle{thmstyleone}%
\theoremstyle{thmstyletwo}%

\theoremstyle{thmstylethree}%

\author*[1]{\fnm{Youval} \sur{Klioui}}\email{y.klioui@tue.nl}

\affil*[1]{\orgdiv{ECE Department}, \orgname{Eindhoven University of Technology}, \orgaddress{\street{Groene Loper 3}, \city{Eindhoven}, \postcode{5612 AE}, \country{Netherlands}}}

\begin{document}

\title[Article Title]{Circulant ADMM-Net for Fast High-resolution DoA Estimation}

\abstract{This paper introduces CADMM-Net and CHADMM-Net, two deep neural networks for direction of arrival (DoA) estimation within the least-absolute shrinkage and selection operator (LASSO) framework. These two networks are based on a structured deep unfolding of the alternating direction method of multipliers (ADMM) algorithm through the use of circulant as well as Hermitian-circulant matrices. Along with a computational complexity of $\mathcal{O}(N\log(N))$ per layer for the inference, where $N$ is the length of the dictionary $\mathbf{A}$, they additionally exhibit a memory footprint of $N$ and approximately half of $N$ for CADMM-Net and CHADMM-Net, respectively, compared with $N^{2}$ for ADMM-Net. Furthermore, these structured networks exhibit a competitive performance against ADMM-Net, THADMM-Net, LISTA, TLISTA, and THLISTA with respect to the detection rate, the angular root-mean square error, and the normalized mean squared error.\blfootnote{The repository for replicating the results reported here can be found at: \codeurl{https://github.com/youvalklioui/cadmmnet}}.}


\maketitle

\section{Introduction}
Fast and high-resolution direction of arrival estimation (DoA) is one key aspect required to enable 
advanced driver-assistance systems \cite{adas1}. Automotive settings add an additional layer of complexity compared with conventional DoA estimation in that the number of snapshots typically available is very limited, with only a single snapshot being the most typical case. This renders the use of subspace-based methods such as the multiple signal classification (MUSIC) \cite{smusic} and estimation of signal parameters via rotational invariant techniques (ESPRIT) \cite{esprit} inadequate unless forward-backward spatial-smoothing (FBSS) \cite{SS} is employed, which reduces the array aperture in addition to constraining the geometry of the full array used since, up to a translation factor, the subarrays themselves must have an identical geometry. This precludes these subspace-based methods from being employed with the typical sparse linear arrays found in an automotive setting such as minimum redundancy arrays \cite{mra}. Maximum likelihood \cite{mle} based methods also prove impractical in this setting as they require an extensive multidimensional grid search, and become even more so for two-dimensional arrays. Gridless methods such as atomic norm minimization (ANM) \cite{ram, numusic} and re-weighted atomic norm minimization (RAM) \cite{anm}, despite their high resolution, also suffer from a high computational cost as they require $M^3$ operations per iteration, where $M$ is the array size, in order to perform a projection onto the space of positive semi-definite matrices per iteration. Similarly, sparse Bayesian learning (SBL) \cite{sbl} is a well-known, high-resolution sparse recovery method that also requires around $M^{3}$ instead of $N^{3}$ operations per iteration through the use of the matrix inversion lemma. The least absolute shrinkage and selection operator (LASSO) \cite{lasso} is another grid-based sparse recovery method that can achieve high resolution beyond the Rayleigh limit while naturally accommodating the single snapshot setting without being restricted to ULAs. While typical solvers used for LASSO such as the iterative shrinkage thresholding algorithm (ISTA) and the alternating direction method of multipliers (ADMM) have a computational complexity per iteration of approximately $MN$ and $M(N+M)$, respectively, they typically require multiple iterations \cite{num_iterations}, often exceeding a hundred, to converge. Additionally, the LASSO, and hence indirectly both ADMM and ISTA, requires properly setting a hyperparameter $\tau$ for the $L_{1}$ norm penalty, where $\tau$ in turn typically depends on the noise variance as well as the sparsity of the signal, both of which are hard to estimate \emph{a priori} using a single snapshot only. Although the Square-root LASSO \cite{sqlasso} addresses the issue of the dependence of $\tau$ over the noise variance, its computational complexity is still considerable, with approximately $N^{3}$ operations per iteration, since it is typically formulated as a second-order conic programming problem and is solved through interior-point methods.  

\par Deep unfolding \cite{lecun} or unrolling was later on introduced to address the high-number-of-iterations aspect of these methods as well as the manual tuning of the LASSO regularization hyperparameter whereby a neural network with an architecture identical to that of the iterative method itself, such as ISTA or ADMM, is trained over a dataset with predefined statistics. Such a procedure typically results in a neural network that can compete with the iterative method being unfolded while having a depth that is considerably smaller than the number of iterations required by the method. Although the number of iterations can be reduced through deep unfolding, for an automotive setting with typically limited compute hardware, the computational cost of doing such an inference for each range-Doppler cell is still considered high. Additionally, each layer of the resulting neural network will require the storage of weight matrices whose dimension will grow with the size of the dictionary used. For instance, Learned-ISTA (LISTA) \cite{lecun}, a deep unrolled  network based on ISTA, requires two weight matrices $\mathbf{W_{1}} \in \mathbb{C}^{N \times N}$ and $\mathbf{W_{2}}\in \mathbb{C}^{N \times M}$ to be  stored for each layer. Structured deep unfolding \cite{tlista} was later on introduced to bring the memory footprint down by restricting the learnable matrices during the training phase to the class of structured ones. For instance, Toeplitz-LISTA \cite{tlista}, or TLISTA, restricts $\mathbf{W_{1}}$ to the set of Toeplitz matrices, thereby reducing the memory footprint from $N^{2}$ to $2N-1$ for $\mathbf{W_{1}}$. The use of Toeplitz matrices also has the added benefit of bringing down the computational cost associated with the multiplication with $\mathbf{W_{1}}$ from $N^{2}$ to $4N\log(2N)$, with the total cost per layer for TLISTA being $MN+4N\log(2N)$ compared to $N^{2}+MN$ for LISTA. Another compact version of LISTA is Trainable-ISTA (TISTA) \cite{tista}. Although TISTA significantly reduces the total number of learnable parameters from $(N^{2}+M)T$ to $T$, where $T$ is the number of layers, this unfolded architecture was derived under the restrictive assumption of a Gaussian dictionary, which does not hold for frequency estimation wherein the dictionary typically consists of a row sub-sampled discrete Fourier transform basis. Taking a similar approach to that used in deriving TLISTA, Toeplitz-Hermitian ADMM-Net \cite{thadmmnet} was subsequently introduced, where, in addition to a Toeplitz constraint on the learnable matrix in ADMM-Net (the network obtained by deep unfolding the ADMM algorithm), a Hermitian as well as a positive-semidefinite constraint is imposed. Although THADMM-Net brings down the total parameter count per layer to $N+2$ while achieving a competitive performance against ADMM-Net, it nevertheless suffers from a high computational complexity of the order of $N^{2}$ operations per layer. 
\par We consider here an extension of this deep-unfolding paradigm that uses learnable constraints and we introduce Circulant ADMM-Net and Circulant-Hermitian ADMM-Net, respectively abbreviated as CADMM-Net and CHADMM-Net. These two deep neural networks exploit the closure of the set of circulant matrices, and of circulant as well as Hermitian matrices, to build compactly structured versions of ADMM-Net that are also computationally efficient as they require  learnable vectors with only $N$ and $\left \lfloor{N/2}\right \rfloor+1$ parameters per layer, respectively, compared with a full weight matrix per layer of $N^{2}$ parameters for ADMM-Net. Moreover, through the use of fast Fourier transforms (FFT) and inverse fast Fourier transforms (IFFT), the computational cost per layer for both CADMM-Net and CHADMM-Net reduces to $2N\log(N)$ and $3N\log(N)$, respectively, compared to $N^{2}$ for ADMM-Net and THADMM-Net. The main contributions of this work can be summarized as follows:
\begin{itemize}
	\item We introduce two neural networks termed CADMM-Net and CHADMM-Net that are based on a structured deep unfolding of ADMM that impose circulant and  Hermitian-circulant, respectively, constraints on the learnable matrices.
	\item We benchmark the performance of the two proposed networks against both ISTA, ADMM as well as ADMM-Net, THADMM-Net, LISTA, TLISTA and THLISTA for both the uniform linear array (ULA) and sparse linear array (SLA) settings. The introduced models exhibit a competitive performance against the baseline deep-unfolded architectures with respect to the detection rate, the angular root mean squared error (RMSE), and the normalized mean squared error of the reconstructed vectors (NMSE), while having a much lower per-layer computational complexity and memory footprint.
\end{itemize}

The rest of the paper is structured as follows. The signal model along with the proposed network architectures are presented in Section \ref{network_architecutre}. Section \ref{experimental_setup} covers in detail the experimental setup as well as the training setup used to benchmark the performance of CADMM-Net and CHADMM-Net. Section \ref{results} presents a performance characterization of the new networks against other baseline deep unfolded architectures with respect to the detection rate, the RMSE, and the NMSE. Section \ref{conclusion} provides a conclusion for this paper.

\par \emph{\underline{Notation:}} $\mathbb{C}$, $\mathbb{R}^{+}$, $\mathbb{N}$ denotes the set of complex, real positive and natural numbers, respectively. Matrices are denoted by bold uppercase letters ($\mathbf{A}$), and vectors by bold lowercase letters ($\mathbf{a}$).  $|z|$ and $\textrm{arg}(z)$ denote the magnitude and argument, respectively, of the complex number $z$. $(.)^{H}$ denotes the Hermitian transpose, $(.)^{\dagger}$ denotes the pseudo-inverse, and $\overline{(.)}$ denotes the complex conjugate. $\textrm{diag}(\mathbf{v})$ denotes the diagonal matrix obtained by placing the elements of $\mathbf{v}$ in a diagonal fashion in the square null matrix. $ \mathcal{CN}(\mathbf{0},\sigma^{2}\mathbf{I})$ denotes the multivariate complex normal distribution with mean $\mathbf{0}$ and covariance matrix $\sigma^{2}\mathbf{I}$.  $\mathbf{X}(:,k)$ denotes the $k$-th column of $\mathbf{X}$. The notation $\mathbf{X} \ge 0$ is used to indicate that $\mathbf{X}$ is a positive semidefinite matrix. $\mathbf{x}\odot\mathbf{y}$ denotes the element-wise product of $\mathbf{x}$ and $\mathbf{y}$. $\mathbf{x}\ast \mathbf{y}$ denotes the discrete convolution of $\mathbf{x}$ with $\mathbf{y}$.  $\mathbf{x}^{-1}$ denotes the element-wise inversion of $\mathbf{x}$, and $\mathbf{1}_{N}$ denotes the ones vector of size $N$. $\left \lfloor{x}\right \rfloor$ denotes the largest integer less than or equal to $x$. $\mathcal{T}^{N}$ denotes the set of complex Toeplitz matrices of size $N$.

\section{Network Architecture}
\label{network_architecutre}

\subsection{Signal Model}
\label{signalmodel}
We consider $K$ targets impinging on a linear array with elements positioned at coordinates $d_{1}, d_{2},\hdots, d_{M}$ with respect to some arbitrary origin. The resulting single-snapshot signal model can be formulated as
\begin{align}
	\mathbf{y}=\sum_{k=1}^{K}c_{k}\mathbf{a}(\theta^{*}_{k})+\mathbf{n}
\end{align}
where $c_{k}\in\mathbb{C}$ is the complex amplitude of the $k$-th target, $\mathbf{n}\sim \mathcal{CN}(\mathbf{0},\sigma^{2}\mathbf{I})$ is complex Gaussian additive noise, and $\mathbf{a}(\theta^{*}_{k})\in\mathbb{C}^{M\times 1}$ is the steering vector defined as
\begin{align}
\mathbf{a}(\theta^{*}_{k})(m)=\exp(-j2\pi\frac{d_{m}}{\lambda}\sin(\theta^{*}_{k})),
\end{align} 
for $m=1, 2, \hdots, M$ . We furthermore assume that the array elements are positioned at integer multiples of a given wavelength unit, i.e. $d_{m}=l_{m}\gamma \lambda$, where $\gamma \in \mathbb{R}^{+}$ and $l_{m}\in\mathbb{N}$. The $k$-th frequency $f^{*}_{k}$ associated with $\theta^{*}_{k}$ is then defined as $f^{*}_{k}=-\gamma\sin(\theta^{*}_{k})$.  With this configuration, we can express the steering vector as $\mathbf{a}(\theta^{*}_{k})(m)=\exp(j2\pi f^{*}_{k} l_{m})$, and  we see that for $\theta_{k}^{*}\in [-\frac{\pi}{2},\frac{\pi}{2}]$, we will have $f_{k}^{*}\in [-\gamma,\gamma]$. Thus, in order to avoid aliasing of the frequencies corresponding to the angles, we must naturally have $\gamma < 1/2$. 

\subsection{LASSO Formulation}
In the LASSO \cite{lasso} context, given the measurement vector $\mathbf{y}$, recovering the frequencies $f^{*}_{k}$ essentially amounts to solving the following optimization problem \cite{lasso}\begin{align}
	\min_{\mathbf{x}} \dfrac{1}{2}||\mathbf{y}-\mathbf{A}\mathbf{x}||_{2}^{2}+\tau||\mathbf{x}||_{1}, \label{lasso1}
\end{align}
where $\tau \in\mathbb{R}^{+}$ is a hyperparameter that balances the data consistency term with the sparsity penalty, and $\mathbf{A}\in \mathbb{C}^{M\times N}$ is the overcomplete dictionary, with $N\gg M$, defined as $\mathbf{A}(m,n) = \exp(j2\pi f_{n} l_{m})$ over the frequency grid points $f_{n}$ for $n=1, 2, \hdots, N$. Estimates of the true frequencies $f^{*}_{k}$ can then be recovered from the support of the solution $\hat{\mathbf{x}}$ to \eqref{lasso1}. 

\subsection{ISTA, LISTA, TLISTA and THLISTA}

The ISTA formulation for solving \eqref{lasso1} is given by the fixed-point iteration
\begin{align}
	\mathbf{x}^{(t)} = S_{\kappa}((\mathbf{I} - \mu \mathbf{A}^{H}\mathbf{A})\,\mathbf{x}^{(t-1)} + \mu \mathbf{A}^{H}\mathbf{y}), \label{ista}
\end{align}
with an initial value $\mathbf{x}^{(0)}$ (typically the zero vector) and a step size defined as $\mu = 1/\sigma_{max}(\mathbf{A})^{2}$, where $\sigma_{max}(\mathbf{A})$ is the largest singular value of the dictionary $\mathbf{A}$, and $S_{\kappa}(z) \triangleq e^{j\arg(z)}\max\bigl(|z| - \kappa, 0\bigr)$ denotes the soft-thresholding operator where the threshold $\kappa$ for ISTA is given by $\kappa = \mu \tau$.

In LISTA, the iterative scheme in \eqref{ista} is replaced by a neural network whose $t$-th layer follows
\begin{align}
	\mathbf{x}^{(t)} = S_{\beta^{(t)}}(\mathbf{W}^{(t)}_{1}\,\mathbf{x}^{(t-1)} + \mathbf{W}^{(t)}_{2}\,\mathbf{y}), \label{lista}
\end{align}
where the parameters $(\mathbf{W}^{(t)}_{1}, \mathbf{W}^{(t)}_{2}, \beta^{(t)})$ are learned. A single LISTA layer entails a per-layer parameter count of $N^2 + MN + 1$ and a computational complexity of $N^2 + MN$. 

TLISTA \cite{tlista} takes advantage of the fact that when the frequencies $f_{n}$ are uniformly spaced (i.e., $f_{n}=f_{1}+(n-1)\Delta f$), the matrix $\mathbf{A}^{H}\mathbf{A}$ becomes Toeplitz, as is $(\mathbf{I} -\mu \mathbf{A}^{H}\mathbf{A})$. In this case, the learnable matrix $\mathbf{W}^{(t)}_{1}$ is constrained to be Toeplitz, i.e. $\mathbf{W}^{(t)}_{1} \in \mathcal{T}^{N}$, and can be fully described by a single vector $\mathbf{w}^{(t)}_{1} \in \mathbb{C}^{2N-1}$. This reduces both the number of parameters and the computational cost per layer to $2N - 1 + MN$ and $MN + 4N\log(2N)$, respectively. Moreover, by imposing an additional Hermitian constraint on $\mathbf{W}^{(t)}_{1}$, i.e. $\mathbf({\mathbf{W}}^{(t)}_{1})^{H} = {\mathbf{W}}^{(t)}_{1}$, one obtains Toeplitz-Hermitian LISTA (THLISTA).
\subsection{ADMM, ADMM-Net and THADMM-Net}

Along with coordinate-descent, ADMM \cite{admm} is widely employed to solve the optimization problem in \eqref{lasso1}. A significant benefit of ADMM over ISTA is that it typically exhibits a higher convergence rate, often requiring far fewer iterations \cite{num_iterations}. The LASSO problem in \eqref{lasso1} is typically reformulated into the following equivalent constrained formulation, making it well-suited for ADMM,
\begin{align}
	&\min_{\mathbf{x}, \mathbf{z}} \frac{1}{2} \|\mathbf{y} - \mathbf{A}\mathbf{x}\|_{2}^{2} + \tau \|\mathbf{z}\|_{1} \\
	&\text{subject to: } \mathbf{x} = \mathbf{z} \nonumber.
\end{align}
The augmented Lagrangian $L_{\rho}(\mathbf{x},\mathbf{z},\mathbf{w})$ of the above problem is then expressed as
\begin{align}
	L_{\rho}(\mathbf{x},\mathbf{z},\mathbf{v})=\frac{1}{2} \|\mathbf{y} - \mathbf{A}\mathbf{x}\|_{2}^{2} + \tau\nonumber \|\mathbf{z}\|_{1}-\frac{\rho}{2}\|\mathbf{v}\|_{2}^{2}+ \frac{\rho}{2}\||\mathbf{x}-\mathbf{z}+\mathbf{v}\|_{2}^{2},
\end{align}
where $\rho \in \mathbb{R}^{+}$ is a hyperparameter, $\mathbf{x}$, $\mathbf{z}$ are the so-called primal variables, and $\mathbf{v}$ is the dual variable. ADMM operates by alternating between a minimization step on the primal variables $(\mathbf{x}, \mathbf{z})$ and a gradient ascent step on the dual variable $\mathbf{v}$
\begin{align}
	\mathbf{x}^{(t)} &= \underset{\mathbf{x}}{\textrm{argmin}} ~	L_{\rho}(\mathbf{x},\mathbf{z}^{(t-1)},\mathbf{v}^{(t-1)})	\\
	\mathbf{z}^{(t)} &= \underset{\mathbf{z}}{\textrm{argmin}} ~	L_{\rho}(\mathbf{x}^{(t)},\mathbf{z},\mathbf{v}^{(t-1)})\\
	\mathbf{v}^{(t)} &= \mathbf{v}^{(t-1)}+\mathbf{x}^{(t)}-\mathbf{z}^{(t)},
\end{align}
which reduces to the following system
\begin{align}
	&\mathbf{x}^{(t)}=(\mathbf{A}^{H}\mathbf{A}+\rho\mathbf{I})^{-1}\big(\mathbf{A}^{H}\mathbf{y}+\rho(\mathbf{z}^{(t-1)}-\mathbf{v}^{(t-1)})\big) \label{x}\\
	&\mathbf{z}^{(t)}=S_{\kappa}(\mathbf{x}^{(t)}+\mathbf{v}^{(t-1)})\label{z}\\
	&\mathbf{v}^{(t)}=\mathbf{x}^{(t)}+\mathbf{v}^{(t-1)}-\mathbf{z}^{(t)}, \label{w}	
\end{align}
where the threshold is given by $\kappa=\rho\tau$ for ADMM, and a typical initialization with $\mathbf{0}$ for $\mathbf{x}^{(0)}$, $\mathbf{z}^{(0)}$, and $\mathbf{v}^{(0)}$. In order to derive the input-output relationships governing ADMM-Net and its structured variants, we will first recast the system above into an equivalent compact form.  Replacing \eqref{z} in \eqref{w}, we obtain
\begin{align}
	\mathbf{v}^{(t)}=\mathbf{x}^{(t)}+\mathbf{v}^{(t-1)}-S_{\kappa}(\mathbf{x}^{(t)}+\mathbf{v}^{(t-1)})\label{w1}.
\end{align}
Plugging \eqref{z} and \eqref{w1} at step $(t-1)$ into \eqref{x}, we get
\begin{align}
	\mathbf{x}^{(t)}&=(\mathbf{A}^{H}\mathbf{A}+\rho\mathbf{I})^{-1}\big(\mathbf{A}^{H}\mathbf{y}+\rho(2S_{\kappa}(\mathbf{x}^{(t-1)}+\mathbf{v}^{(t-2)}) -\mathbf{x}^{(t-1)}-\mathbf{v}^{(t-2)})\big) \label{x2}.
 \end{align}
Adding \eqref{w1} at step $(t-1)$ to \eqref{x2}, defining the auxiliary variable $\mathbf{u}^{(t)}=\mathbf{x}^{(t)}+\mathbf{v}^{(t-1)}$, and setting $\mathbf{y_F}=\mathbf{A}^{H}\mathbf{y}$, we obtain
\begin{align}
	\mathbf{u}^{(t)}=(\mathbf{A}^{H}\mathbf{A}+\rho\mathbf{I})^{-1}(\mathbf{y_{F}}+\rho(2S_{\kappa}(\mathbf{u}^{(t-1)}) 
	-\mathbf{u}^{(t-1)}))+\mathbf{u}^{(t-1)}-S_{\kappa}(\mathbf{u}^{(t-1)}). \label{u}
\end{align}
For a large enough number of iterations $T$, we have the approximation $\mathbf{u}^{(T)}\approx \mathbf{u}^{(T-1)}$, along with this resulting approximation
\begin{align}
	&\mathbf{x}^{(T)}=\mathbf{u}^{(T)}-\mathbf{v}^{(T-1)}=\mathbf{u}^{(T)}-\big(\mathbf{u}^{(T-1)}-S_{\kappa}(\mathbf{u}^{(T-1)})\big)  \nonumber \\ &\approx \mathbf{u}^{(T)}-\big(\mathbf{u}^{(T)}-S_{\kappa}(\mathbf{u}^{(T)})\big)  \approx S_{\kappa}(\mathbf{u}^{(T)}). \label{x3}
\end{align}
The compact approximate version of ADMM, summarized in Algorithm \ref{alg1}, serves as the basis for ADMM-Net and its structured variants as it allows us to recover $\mathbf{x}^{(T)}$ directly from $\mathbf{u}^{(T)}$  without the need to have $\mathbf{u}^{(T-1)}$ stored in memory. 
\begin{algorithm}[h!]
	\caption{Compact Approximate ADMM}
	Input: $\rho, \tau>0$, $\mathbf{u^{0}}\in\mathbb{C}^{N\times1}$, $\mathbf{y}\in\mathbb{C}^{M\times1}$, $\mathbf{A}\in\mathbb{C}^{M\times N}$, $T$\;
	$\mathbf{y_{F}}=\mathbf{A}^{H}\mathbf{y}, \kappa = \rho\tau$\;
	$\mathbf{Q}=(\mathbf{A}^{H}\mathbf{A}+\rho\mathbf{I})^{-1}$\;
	\For{$t := 1$ to $T$}
	{
			$\mathbf{u}^{(t)}=\mathbf{Q}\big(\mathbf{y_{F}}+\rho^{(t)}(2S_{\kappa}(\mathbf{u}^{(t-1)})-  
			\mathbf{u}^{(t-1)})\big)+ \mathbf{u}^{(t-1)}-S_{\kappa}(\mathbf{u}^{(t-1)}) $ \;
	}
	Output: $\mathbf{x}^{(T)}= S_{\kappa}(\mathbf{u}^{(T)})$\;
	\label{alg1}
\end{algorithm}
\noindent The input-output relationship of the $t$-th layer of ADMM-Net is then given by
\begin{align}
	\mathbf{u}^{(t)}=(\mathbf{W}^{(t)}+\rho^{(t)}\mathbf{I})^{\dagger}\big(\mathbf{y_{F}}+\rho^{(t)}(2S_{\beta^{(t)}}(\mathbf{u}^{(t-1)}) 
	-\mathbf{u}^{(t-1)})\big)+\mathbf{u}^{(t-1)}-S_{\beta^{(t)}}(\mathbf{u}^{(t-1)}) \label{admmnet},
\end{align}
along with only a learnable soft-thresholding operation performed on the output layer
\begin{align}
	\mathbf{x}^{(T)}=S_{\beta^{(T+1)}}(\mathbf{u}^{(T)}).
\end{align}
The learnable parameters of the $t$-th layer for ADMM-Net are thus $(\mathbf{W}^{(t)},\beta^{(t)},\rho^{(t)})$, with a parameter count of $N^{2}+2$ per layer, and a computational cost of $N^{3}+N^{2}$ operations during the training stage, and $N^{2}$ for the inference stage. Similarly to THLISTA, an equivalent version based on ADMM-Net called THADMM-Net \cite{thadmmnet} can also be obtained whereby the learnable matrix $\mathbf{W}^{(t)}$ is restricted to the set of Toeplitz-Hermitian matrices in addition to a positive semidefinite constraint. These three constraints can be simultaneously imposed on $\mathbf{W}^{(t)}$ in \eqref{admmnet} through the following parameterization \cite{thadmmnet}
\begin{align}
	\mathbf{W}^{(t)} \triangleq \mathbf{W}_{TH}^{(t)} + \max(-\lambda_{min}(\mathbf{W}_{TH}^{(t)}), 0)\mathbf{I},  \label{thadmmnet}
\end{align}
where $\lambda_{min}$ denotes the smallest eigenvalue of $\mathbf{W}_{TH}^{(t)}$, and $\mathbf{W}^{(t)}\in \mathcal{T}$ in addition to the Hermitian constraint $\mathbf{W}_{TH}^{(t)} = (\mathbf{W}_{TH}^{(t)})^{H}$. With these constraints, THADMM-Net requires $N+2$ per layer whereas the computational complexity remains similar to that of ADMM-Net.
\subsection{CADMM-Net and CHADMM-Net}
 
 Although THADMM-Net is more efficient with respect to the parameter count when compared to ADMM-Net, it nevertheless suffers from high computational cost. Thus an efficient architecture that reduces both the number of parameters as well as the computational complexity without sacrificing performance is highly desirable. The following proposition motivates the use of a circulant constraint on the learnable matrices in the classical ADMM-Net architecture.\\

 \textbf{Proposition:} Let an arbitraty array be specified by its indices $\{p_{m}\}_{m=1}^{M}$ where $p_{m} = \gamma \l_{m}$ for $\{l_{m}\}_{m=1}^{M}\subset \mathbb{Z}$ and $0<\gamma \le 1/2$. Assume a uniform frequency grid given by $f_n=-\gamma +2\gamma(n-1)/{N}$, for $n=1, 2, \hdots, N$. Then the matrix $(\mathbf{A}^{H}\mathbf{A}+\rho\mathbf{I})^{-1}$ is circulant for any sequence of array linear indices $\{l_{m}\}_{m=1}^{M} \subset \mathbb{Z}$  if and only if $\gamma=1/2$. \vspace{0.5cm}

\emph{Proof:} We will make use of the following definition for circulant matrices: $\mathbf{C}\in \mathbb{C}^{N\times N}$ is circulant if $\mathbf{C}$ is Toeplitz, i.e.  $\mathbf{C}(n,p)=\mathbf{C}(n-p)$, and $\mathbf{C}(n,p)=\mathbf{C}(n-p+N)$ for $n-p< 0$.  For the reverse implication, with the definition $\mathbf{B}=\mathbf{A}^{H}\mathbf{A}$, we have
\begin{align}
	\mathbf{B}(n,p)&=\sum_{m=1}^{M} \exp(j2\pi(f_{p}-f_{n})l_{m}) \nonumber\\&=
	\sum_{m=1}^{M} \exp(j4\pi\gamma\frac{(p-n)}{N}l_{m})
	\nonumber\\&=
	\sum_{m=1}^{M} \exp(j2\pi\frac{(p-n)}{N}l_{m})=\mathbf{B}(n-p).
\end{align}
Thus a uniformity constraint on the grid enforces a Toeplitz constraint on $\mathbf{B}$.  We additionally have
\begin{align}
	\mathbf{B}(n-p+N)
	&=\sum_{m=1}^{M}\exp(j2\pi l_{m}) \exp(j2\pi\frac{(p-n)}{N}l_{m}) \nonumber \\
	&=\sum_{m=1}^{M} \exp(j2\pi\frac{(p-n)}{N}l_{m})=\mathbf{B}(n-p). \label{circ2}
\end{align}
Thus $\mathbf{B}$ is circulant. Using the closure under addition and inversion of circulants concludes the proof for the reverse. For the forward implication, we prove the contrapositive. Namely if $\gamma \neq 1/2$, then there exists at least one sub-sequence $\mathcal{L}=\{l_{m}\}_{m=1}^{M} \subset \mathbb{Z}$ such that $(\mathbf{A}^{H}\mathbf{A}+\rho\mathbf{I})^{-1}$ is not circulant. For $M=1$ and $\mathcal{L} = \{1\}$, we have
\begin{align}
\mathbf{B}(n-p+N)-\mathbf{B}(n-p) = (1-\exp(j4\pi \gamma)) \exp(j4\pi\gamma\frac{p-n}{N}).
\end{align}
For $0<\gamma < 1/2$ and $0<|n-p|<N$, we have $(1-\exp(j4\pi \gamma)) \neq 0$ and $\exp(j4\pi\gamma\frac{p-n}{N}) \neq 0$, thus $\mathbf{B}(n-p+N)-\mathbf{B}(n-p) \neq 0$, from which $\mathbf{B}$, and hence $(\mathbf{A}^{H}\mathbf{A}+\rho\mathbf{I})^{-1}$, is not circulant.$\blacksquare$

\par
\begin{sidewaysfigure}[htbp]
    \centering
    \includegraphics[height=3.8cm,width=19cm]{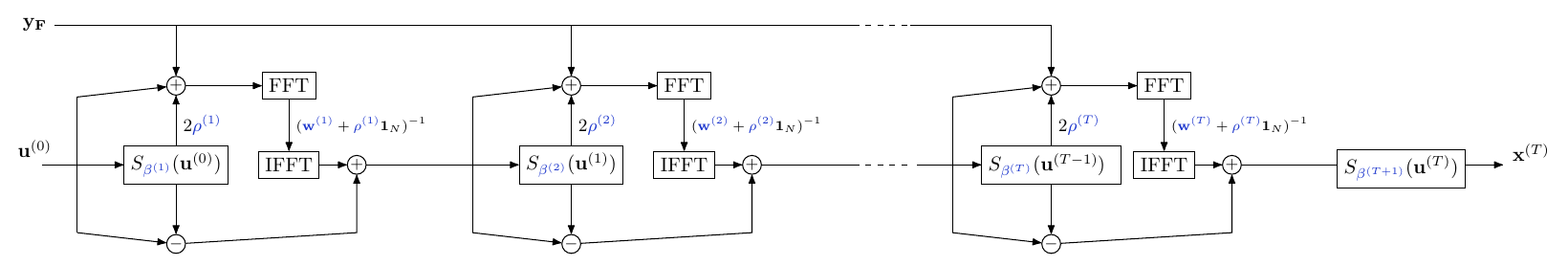}
    \caption{CADMM-Net network with $T$ layers. The learnable parameters are highlighted in blue.}
    \label{network}
    \vskip 2cm
    \centering
    % Explicit threeparttable removed to prevent the nested loop crash
    \caption{Summary of CADMM-Net and CHADMM-Net characteristics.}
    \label{table}
    \begin{tabular}{c|c|c|c|c|c|}
    \hhline{~|-----}
    & ADMM-Net & THADMM-Net & CADMM-Net & CHADMM-Net & TLISTA \\ \hline
    \multicolumn{1}{|c|}{\makecell{Parameters\\per layer}} & $N^{2}+2$ & $N+2$ & $N+2$ & $\left\lfloor N/2 \right\rfloor+3$ & $2N+MN$ \\ \hline
    \multicolumn{1}{|c|}{\makecell{Computational\\ complexity per layer}} & \makecell{Training: $N^{3}+N^{2}$\\ Inference: $N^{2}$} & \makecell{Training: $N^{3}+N^{2}$\\ Inference: $N^{2}$} & $2N\log(N)$ & $3N\log(N)$ & \makecell{4$N\log(2N)$\\+ $MN$} \\ \hline
    \end{tabular}%

\end{sidewaysfigure}

 It is worth mentioning that with the requirement of a uniform grid only, $(\mathbf{A}^{H}\mathbf{A}+\rho\mathbf{I})^{-1}$ will in general be a centro-Hermitian \cite{centrohermitian} matrix that exhibits a conjugate symmetry with respect to its center. Similarly to circulants, such matrices are also closed under addition, multiplication and inversion and can be characterized by approximately $\lfloor{N^{2}/4\rfloor}$ elements. As mentioned in \ref{signalmodel}, in order to guarantee the non-aliasing of the frequencies corresponding to the angles of arrival, we must have $0<\gamma<1/2$. For the specific case of $\gamma=1/2$, the two angles $\theta_{1}=-\pi/2$ and $\theta_{2}=\pi/2$ would be indistinguishable  as they would both map to the same frequency $f=-1/2$. This inconvenience may be alleviated by the fact that the radiation pattern for the antennas, very often planar, employed in automotive arrays rarely cover the full field of view from $-\pi/2$ to $\pi/2$. We will thus henceforth  assume a uniform frequency grid and set $\gamma=1/2$. Using the well-known eigendecomposition of circulant matrices, we can then write
 \begin{align}
(\mathbf{A}^{H}\mathbf{A}+\rho\mathbf{I})^{-1}&=(\mathbf{F}^{-1}\textrm{diag}(\mathbf{F}\mathbf{b})\mathbf{F}+\rho\mathbf{F}^{-1}\mathbf{F})^{-1} =(\mathbf{F}^{-1}(\textrm{diag}(\mathbf{F}\mathbf{b})+\rho\mathbf{I})\mathbf{F})^{-1} \nonumber
\\&=\mathbf{F}^{-1}(\textrm{diag}(\mathbf{F}\mathbf{b})+\rho\mathbf{I})^{-1}\mathbf{F}, \end{align}
where $\mathbf{F}$ and $\mathbf{F}^{-1}$ are the discrete Fourier transform (DFT) and inverse discrete Fourier transform (IDFT) matrices, respectively, and $\mathbf{b}$ is the first column of $\mathbf{B}=\mathbf{A}^{H}\mathbf{A}$. The input-output relationship of the $t$-th layer of CADMM-Net can then be stated as
\begin{align}
	\mathbf{u}^{(t)}&=\mathbf{F}^{-1}(\mathbf{w}^{(t)}+\rho^{(t)}\mathbf{1}_{N})^{-1}\odot\mathbf{F}\big(\mathbf{y_{F}} +\rho^{(t)}(2S_{\beta^{(t)}}(\mathbf{u}^{(t-1)})-\mathbf{u}^{(t-1)})\big) \nonumber \\
	&+\mathbf{u}^{(t-1)}-S_{\beta^{(t)}}(\mathbf{u}^{(t-1)}),
\end{align}
where the learnable parameters are ($\mathbf{w}^{(t)},\rho^{(t)},\beta^{(t)})$. At the output of the network, $\mathbf{x}^{(T)}$ is recovered in the last layer through a learnable thresholding operation
\begin{align}
	\mathbf{x}^{(T)}=S_{\beta^{(T+1)}}(\mathbf{u}^{(T)}).
\end{align}
The corresponding parameter count per layer is thus $N+2$, and the computational cost is $2N\textrm{log}(N)$ operations for both training and inference, mainly dominated by the FFT and IFFT operations. The corresponding neural network architecture of a $T$-layer CADMM-Net network is shown in Fig. \ref{network}. In addition to being circulant, $\mathbf{B}=\mathbf{A}^{H}\mathbf{A}$ is also Hermitian, and so will be $(\mathbf{A}^{H}\mathbf{A}+\rho\mathbf{I})^{-1}$ since the Hermitian and inverse operators commute. A circulant Hermitian matrix is the complex extension of the circulant symmetric matrix and is highly structured as it can be fully characterized by only $\left \lfloor{N/2}\right \rfloor+1$ complex scalars. The vector $\mathbf{c}$ parameterizing a circulant Hermitian matrix $\mathbf{C}$ has the following form,
\begin{align}
	\mathbf{c}=[c_{1}, c_{2}, \hdots, c_{\left \lfloor{N/2}\right \rfloor+1}, \overline{c}_{N-\left \lfloor{N/2}\right \rfloor},\overline{c}_{N-\left \lfloor{N/2}\right \rfloor-1},\hdots, \overline{c}_{2}  ].
\end{align}
The input-output relationship for CHADMM-Net is then given by
\begin{align}
	\mathbf{u}^{(t)}&=\mathbf{F}^{-1}(\mathbf{F}\mathbf{w}^{(t)}+\rho^{(t)}\mathbf{1}_{N})^{-1}\odot\mathbf{F}\big(\mathbf{y_{F}} +\rho^{(t)}( 
	 2S_{\beta^{(t)}}(\mathbf{u}^{(t-1)})-\mathbf{u}^{(t-1)})\big) \nonumber \\ &+\mathbf{u}^{(t-1)}-S_{\beta^{(t)}}(\mathbf{u}^{(t-1)}),
\end{align}
where the learnable parameters are ($\mathbf{w}^{(t)},\rho^{(t)},\beta^{(t)})$ and $\mathbf{w}^{(t)} \in \mathbb{C}^{N\times1}$ satisfies the constraint
\begin{align}
	\mathbf{w}^{(t)}(n)=\overline{\mathbf{w}^{(t)}}(N+2-n), \quad n=\left \lfloor{N/2}\right \rfloor+2,\hdots,N. \label{ch}
\end{align} 

\par It is worth mentioning that with the requirement of a uniform grid only, $(\mathbf{A}^{H}\mathbf{A}+\rho\mathbf{I})^{-1}$ will in general be a centro-Hermitian \cite{centrohermitian} matrix that exhibits a conjugate symmetry with respect to its center. Similarly to circulants, such matrices are also closed under addition, multiplication and inversion and can be characterized by approximately $\lfloor{N^{2}/4\rfloor}$ elements. As mentioned in \ref{signalmodel}, in order to guarantee the non-aliasing of the frequencies corresponding to the angles of arrival, we must have $0<\gamma<1/2$. For the specific case of $\gamma=1/2$, the two angles $\theta_{1}=-\pi/2$ and $\theta_{2}=\pi/2$ would be indistinguishable  as they would both map to the same frequency $f=-1/2$. This inconvenience may be alleviated by the fact that the radiation pattern for the antennas, very often planar, employed in automotive arrays rarely cover the full field of view from $-\pi/2$ to $\pi/2$. We will thus henceforth  assume a uniform frequency grid and set $\gamma=1/2$. Using the well-known eigendecomposition of circulant matrices, we can then write
 \begin{align}
(\mathbf{A}^{H}\mathbf{A}+\rho\mathbf{I})^{-1}&=(\mathbf{F}^{-1}\textrm{diag}(\mathbf{F}\mathbf{b})\mathbf{F}+\rho\mathbf{F}^{-1}\mathbf{F})^{-1} =(\mathbf{F}^{-1}(\textrm{diag}(\mathbf{F}\mathbf{b})+\rho\mathbf{I})\mathbf{F})^{-1} \nonumber
\\&=\mathbf{F}^{-1}(\textrm{diag}(\mathbf{F}\mathbf{b})+\rho\mathbf{I})^{-1}\mathbf{F}, \end{align}
where $\mathbf{F}$ and $\mathbf{F}^{-1}$ are the discrete Fourier transform (DFT) and inverse discrete Fourier transform (IDFT) matrices, respectively, and $\mathbf{b}$ is the first column of $\mathbf{B}=\mathbf{A}^{H}\mathbf{A}$. The input-output relationship of the $t$-th layer of CADMM-Net can then be stated as
\begin{align}
	\mathbf{u}^{(t)}&=\mathbf{F}^{-1}(\mathbf{w}^{(t)}+\rho^{(t)}\mathbf{1}_{N})^{-1}\odot\mathbf{F}\big(\mathbf{y_{F}} +\rho^{(t)}(2S_{\beta^{(t)}}(\mathbf{u}^{(t-1)})-\mathbf{u}^{(t-1)})\big) \nonumber \\
	&+\mathbf{u}^{(t-1)}-S_{\beta^{(t)}}(\mathbf{u}^{(t-1)}),
\end{align}
where the learnable parameters are ($\mathbf{w}^{(t)},\rho^{(t)},\beta^{(t)})$. At the output of the network, $\mathbf{x}^{(T)}$ is recovered in the last layer through a learnable thresholding operation
\begin{align}
	\mathbf{x}^{(T)}=S_{\beta^{(T+1)}}(\mathbf{u}^{(T)}).
\end{align}
The corresponding parameter count per layer is thus $N+2$, and the computational cost is $2N\textrm{log}(N)$ operations for both training and inference, mainly dominated by the FFT and IFFT operations. The corresponding neural network architecture of a $T$-layer CADMM-Net network is shown in Fig. \ref{network}. In addition to being circulant, $\mathbf{B}=\mathbf{A}^{H}\mathbf{A}$ is also Hermitian, and so will be $(\mathbf{A}^{H}\mathbf{A}+\rho\mathbf{I})^{-1}$ since the Hermitian and inverse operators commute. A circulant Hermitian matrix is the complex extension of the circulant symmetric matrix and is highly structured as it can be fully characterized by only $\left \lfloor{N/2}\right \rfloor+1$ complex scalars. The vector $\mathbf{c}$ parameterizing a circulant Hermitian matrix $\mathbf{C}$ has the following form,
\begin{align}
	\mathbf{c}=[c_{1}, c_{2}, \hdots, c_{\left \lfloor{N/2}\right \rfloor+1}, \overline{c}_{N-\left \lfloor{N/2}\right \rfloor},\overline{c}_{N-\left \lfloor{N/2}\right \rfloor-1},\hdots, \overline{c}_{2}  ].
\end{align}
The input-output relationship for CHADMM-Net is then given by
\begin{align}
	\mathbf{u}^{(t)}&=\mathbf{F}^{-1}(\mathbf{F}\mathbf{w}^{(t)}+\rho^{(t)}\mathbf{1}_{N})^{-1}\odot\mathbf{F}\big(\mathbf{y_{F}} +\rho^{(t)}( 
	 2S_{\beta^{(t)}}(\mathbf{u}^{(t-1)})-\mathbf{u}^{(t-1)})\big) \nonumber \\ &+\mathbf{u}^{(t-1)}-S_{\beta^{(t)}}(\mathbf{u}^{(t-1)}),
\end{align}
where the learnable parameters are ($\mathbf{w}^{(t)},\rho^{(t)},\beta^{(t)})$ and $\mathbf{w}^{(t)} \in \mathbb{C}^{N\times1}$ satisfies the constraint
\begin{align}
	\mathbf{w}^{(t)}(n)=\overline{\mathbf{w}^{(t)}}(N+2-n), \quad n=\left \lfloor{N/2}\right \rfloor+2,\hdots,N. \label{ch}
\end{align} 
It worth noting here that CHADMM-Net trades off memory for computational complexity compared to CADMM-Net since the number of parameters per layer is halved but the computational complexity increases to around $3N\log(N)$ operations per layer.  During the training phase of CHADMM-Net, the constraint in \eqref{ch} is applied immediately after  $\mathbf{w}^{(t)}$ is updated through gradient descent. Table \ref{table} provides a summary of the main characteristics of CADMM-Net and CHADMM-Net along with other deep unfolded architectures. As will be shown in the experimental section,  despite having almost half the number of parameters per layer, CHADMM-Net shows a very close performance to that of CADMM-Net.
% --- The Figure ---
\section{Experimental Setup}
\label{experimental_setup}
\subsection{Datasets}
\subsection{Datasets}
For the training setup of CADMM-Net and CHADMM-Net, we fix the dictionary size to $N=256$ and set $\gamma=1/2$. Experiments are conducted in both ULA and SLA scenarios using an array with $M=30$ elements. In the SLA case, $30$ elements are randomly subsampled from a ULA that spans a total aperture of $60\lambda/2$. For each setting, a training set of $10^{5}$ measurement vectors and a validation set of $2\times10^{4}$ measurement vectors are generated together with their corresponding ground truth vectors. The number of targets $K$ is selected uniformly at random from between $1$ and $8$. Afterwards, $K$ frequencies $\{f_{k}^{*}\}_{k=1}^{K}$ are drawn from the interval $[-1/2,1/2[$ while ensuring that the difference between any two frequencies is at least $1/M$.\footnote{The reason we generate the normalized frequencies directly, instead of generating $K$ angles $\{\theta_{k}^{*}\}_{k=1}^{K}$ and then converting them into their corresponding normalized frequency through $f_{k}^{*} = 1/2 \sin(\theta_{k}^{*})$, is that the resolvability of signals is fundamentally determined by how far apart the normalized frequencies \cite{super-resolution} are, as opposed to how far apart the angles are.} Subsequently, for generating the ground truth vectors, each frequency $f_{k}^{*}$ is mapped to its nearest grid point $f_{k}$. For every $f_{k}$, an amplitude $x_{k}$ is generated with $|x_{k}|\sim \mathcal{U}(0,1)$ and $\textrm{arg}(x_{k})\sim \mathcal{U}(0,2\pi)$, thereby constructing the ground truth vector $\mathbf{x}$. Finally, the signal-to-noise ratio (SNR) is fixed at 15 dB, and the measurement vector $\mathbf{y}$ corresponding to $\mathbf{x}$ is produced as
\begin{align}
	\mathbf{y}= \sum_{k=1}^{K}x_{k}\mathbf{a}(f^{*}_{k})+\mathbf{n},
\end{align}
with $\mathbf{n}\sim \mathcal{CN}(\mathbf{0},\sigma^{2}\mathbf{I})$, where $\sigma^{2}=\frac{1}{M}10^{-\frac{\textrm{SNR}}{10}}\Bigl\|\sum_{k=1}^{K}x_{k}\mathbf{a}(f^{*}_{k})\Bigr\|_{2}^{2}.$
For the performance evaluation in terms of the NMSE, RMSE, and detection rate, we decrease the minimum frequency separation constraint from $1/M$ to $1/(3M)$ and generate a test set consisting of $10^{3}$ measurement vectors, together with their corresponding ground truth, for each SNR level ranging from 0 dB to 35 dB in increments of 5 dB.
\subsection{Training Setup }
We compare CADMM-Net and CHADMM-Net against ADMM-Net, THADMM-Net, LISTA, TLISTA and THLISTA as well as a 30 iteration of ADMM and ISTA. We fix the number of layers $T$ for all the networks to $30$. Table \ref{table3} summarizes the model parameter count for each network for this setup in addition to the number of Complex Floating Point Operations (CFLOPS) required for a single inference. For the initialization, all the weight matrices and vectors in all the layers of all the networks are initialized with their respective value in the corresponding iterative method. For instance, for CADMM-Net, $\mathbf{w}^{(t)}$ is initialized with $\mathbf{Fb}$, where $\mathbf{b}$ is the first column of $\mathbf{A}^{H}\mathbf{A}$, for all layers $t=1, 2, \hdots 30$. Table \ref{table2} provides a summary of the initialization and constraints for the weights of the networks used. Additionally, $\beta^{(t)}$ is initialized with $10^{-1}$ for all networks and $\rho^{(t)}$ is initialized with 1 for ADMM-Net and its structured variants.  We use the Adam \cite{adam} optimizer with a learning rate of $10^{-4}$ along with a batch size of 2048. All the models are trained on an NVIDIA A100 GPU using the PyTorch 2.5.1 + CUDA 12.4 (cuDNN 9.1) library.
\subsection{Support-aware Loss Function}
\begin{figure}
	\centering
	\includegraphics[width=5cm]{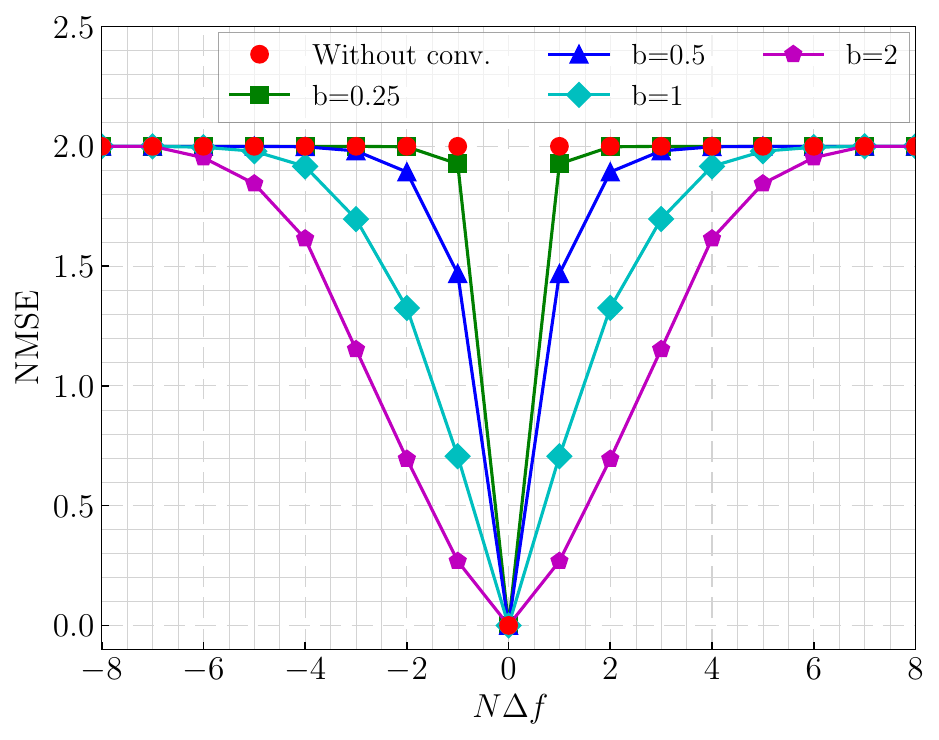}
	\caption{NMSE between a single-component sparse vector and a support-shifted version, both of which are convolved with a Laplacian kernel.}
	\label{fig:conv}
\end{figure}

The standard NMSE loss in the way it is typically used between the output of the neural network $\mathbf{x}^{(T)}$ and the ground truth vector $\mathbf{x}$,
\begin{align}
	\textrm{NMSE}(\mathbf{x}^{(T)},\mathbf{x})=\frac{||\mathbf{x}^{(T)}-\mathbf{x}||_{2}^{2}}{||\mathbf{x}||^{2}},
\end{align}
\begin{sidewaystable*}
    \centering
    \begin{threeparttable}
    \caption{Constraints and initialization of the learnable weights.}
    \label{table2}
    {\renewcommand{\arraystretch}{1.2}
    \begin{tabular}{c|c|c|c|c|c|}
    \hhline{~|-----}
    & ADMM-Net & THADMM-Net & CADMM-Net & CHADMM-Net & TLISTA \\ \hline
    \multicolumn{1}{|c|}{Weights} & $\mathbf{W}^{(t)}$ & $\mathbf{W}^{(t)}$ & $\mathbf{w}^{(t)}$ &\rule{0pt}{14pt}  $\mathbf{w}^{(t)}$ & $(\mathbf{W}_{1}^{(t)},\mathbf{W}_{2}^{(t)})$ \\ \hline
    \multicolumn{1}{|c|}{Constraint} & $\mathbf{W}^{(t)}\in \mathbb{C}^{N\times N}$ & \makecell{\rule{0pt}{14pt}$\mathbf{W}^{(t)} \in \mathcal{T}^{N}$ \\ $(\mathbf{W}^{(t)})^{H} = \mathbf{W}^{(t)}$ \\ $\mathbf{W}^{(t)} \ge 0$} & $\mathbf{w}^{(t)}\in \mathbb{C}^{N\times 1}$ & \makecell{\rule{0pt}{14pt}$\mathbf{w}^{(t)}(n)=\overline{\mathbf{w}}^{(t)}(N+2-n)$ \\ $n=\left\lfloor N/2 \right\rfloor+2,\dots,N$} & \makecell{\rule{0pt}{14pt}$\mathbf{W}_{1}^{(t)} \in \mathcal{T}^{N}$ \\ $\mathbf{W}_{2}^{(t)}\in \mathbb{C}^{M\times N}$} \\ \hline
    \multicolumn{1}{|c|}{Initialization} & $\mathbf{W}^{(t)}=\mathbf{A}^{H}\mathbf{A}$ & $\mathbf{W}^{(t)}=\mathbf{A}^{H}\mathbf{A}$ & $\mathbf{w}^{(t)}=\mathbf{F}\mathbf{b}$ & $\mathbf{w}^{(t)}=\mathbf{b}$ & \makecell{\rule{0pt}{14pt}$\mathbf{W}_{1}^{(t)}=\mathbf{A}^{H}\mathbf{A}$ \\ $\mathbf{W}_{2}^{(t)}= \mu \mathbf{A}^{H}$} \\ \hline
    \end{tabular}
    }
    \end{threeparttable}
    \vskip 1.5cm

    \begin{threeparttable}
    \caption{Model parameter count and total CFLOPs for an inference for the experimental setup.}
    \label{table3}
    \begin{tabular}{c|c|c|c|c|c|}
    \hhline{~|-----}
    & ADMM-Net & THADMM-Net & CADMM-Net & CHADMM-Net & TLISTA \\ \hline
    \multicolumn{1}{|c|}{\makecell{Model \\ Parameters}} & 1966140 & 7740 & 7740 & $\mathbf{3930}$ & 245760 \\ \hline
    \multicolumn{1}{|c|}{\makecell{Inference \\ CFLOPS}} & 1966080 & 1966080 & $\mathbf{122880}$ & 184320 & 506880 \\ \hline
    \end{tabular}
\end{threeparttable}
\end{sidewaystable*}
is not a smooth function with respect to the support itself. For instance, let us consider a sparse ground truth vector consisting of a single non-zero component, $\mathbf{x}(n)=x_{0}\delta(f_{n}-f_{n_{0}})$. Assuming an output from the neural net of a similar form but with a shifted support, $\mathbf{x}^{(T)}(n)=x_{1}\delta(f_{n}-f_{n_{1}})$, the resulting NMSE as a function of the mismatch with respect to the frequency support, $\Delta f=|f_{n_{0}}-f_{n_{1}}|$, would assume the following form,
\begin{align}
	\textrm{NMSE}(\Delta f)= \begin{cases} 
		|x_{1}-x_{0}|^{2}/|x_{0}|^{2} & \text{if } \Delta f = 0 \\
		(|x_{1}|^{2}+|x_{0}|^{2})/|x_{0}|^{2} & \text{if } \Delta f \neq 0.
	\end{cases}
\end{align}
Thus, when there is a mismatch, this loss ascribes the same constant value, regardless of how large or small this mismatch can be. Therefore, before computing the loss, both $\mathbf{x}^{(T)}$ and $\mathbf{x}$ are convolved with a zero-mean Laplacian kernel $\mathbf{g}$ with a scale parameter $b>0$
\begin{align}
	\mathbf{g}(n)=\exp(-|n|/b), \quad n=-\left \lfloor{N/2}\right \rfloor, \hdots,\left \lfloor{N/2}\right \rfloor,
\end{align} 
we afterwards use $\textrm{NMSE}(\mathbf{x}^{(T)} \ast \mathbf{g}, \mathbf{x}\ast \mathbf{g} )$ as the loss function. Using the same previous example of a sparse vector with a single component.  Fig. \ref{fig:conv} shows the resulting loss with this method. Naturally, the increase in sensitivity of the loss function with respect to $\Delta f$ for higher values of $b$ is simultaneous with the risk of lower amplitude targets being submerged by the tails of the stronger ones after the convolution operation. We thus settle for $b=1/2$ as a middle ground for our loss function. It is worth mentioning that another loss function that quantifies support mismatch is the Sinkhorn distance \cite{sinkhorn1}. However, this loss function requires multiple iterations to converge, and thus effectively extends the depth of the network during the training phase since the full computational graph \cite{autograd} needs to be stored during the forward pass. It additionally suffers from numerical instability issues, and although there is a log-stabilized version \cite{sinkhorn2}, it still adds a considerable overhead for both memory and compute and proved to be too impractical to use in our experiments. 

\subsection{Performance Metrics}
To evaluate the performance of CADMM-Net and CHADMM-Net against the other networks, we employ the detection rate and the root mean-squared error (RMSE). First, using either a neural network or an iterative technique, we obtain an estimate $\mathbf{\hat{x}}$ of the ground-truth vector $\mathbf{x}$ corresponding to a measurement vector $\mathbf{y}$. We then perform peak detection on the amplitude spectrum to derive a new spectrum $\mathbf{\hat{x}}_{pk}$ that retains only the peak values from $|\mathbf{\hat{x}}|$. For a target in the ground-truth spectrum $|\mathbf{x}|$ that appears at a frequency bin indexed by $r$ on the uniform grid, we identify the set of indices
\begin{align}
	\mathcal{I}_{r}=\{\hat{r}_{1}, \hat{r}_{2}, \ldots, \hat{r}_{L(r)}\},
\end{align}
comprising the estimated support values from $|\mathbf{\hat{x}}|$ that satisfy
\begin{align}
	|r-\hat{r}_{l}| \le \delta_{1}, \qquad l = 1, 2, \ldots, L(r).
\end{align}
Next, from $\mathcal{I}_{r}$ we keep only those indices forming the subset $\mathcal{J}_{r}\subseteq \mathcal{I}_{r}$ where the ratio between the amplitude of the estimated spectrum and the ground-truth amplitude exceeds a predefined threshold $0<\delta_{2}\le 1$, i.e.,
\begin{align}
	\dfrac{\mathbf{\hat{x}}_{pk}(j)}{|\mathbf{x}(r)|} \ge \delta_{2}, \qquad j \in \mathcal{J}_{r}.
\end{align}
If $\mathcal{J}_{r}$ is non-empty, then the recovery corresponding to the target at frequency bin $r$ is deemed successful. This process is repeated for all $K$ targets present in a ground-truth vector $\mathbf{x}$. Defining the indicator function as $\mathds{1}(\phi)= 1$ if $\phi \neq \emptyset$ and $0$ otherwise, the detection rate $P_{d}$ is given by
\begin{align}
	P_{d} =\dfrac{\sum_{k=1}^{K}\mathds{1}(\mathcal{J}_{r(k)})}{K}.
\end{align}
For each SNR level, we report the mean detection rate computed over a test set of 1000 measurement vectors. For the RMSE computation, consider the $k$-th ground-truth target located at the frequency bin $r(k)$. We first determine the index $\tilde{j}(k)$ in $\mathcal{J}_{r(k)}$ that is closest to $r(k)$
\begin{align}
	\tilde{j}(k)=\argmin \limits_{j\in \mathcal{J}_{r(k)}}|j-r(k)|.
\end{align}
Next, denoting by $\theta_{n}=\sin^{-1}(f_{n})/\gamma$ the angle corresponding to the $n$-th frequency bin, the mean squared angular error $E$ corresponding to a single test vector is then calculated as 
\begin{align}
	E=\dfrac{1}{N_{d}}\sum_{s=1}^{N_{d}}\Bigl(\theta_{r(k_{s})}-\theta_{\tilde{j}(k_{s})}\Bigr)^{2},
\end{align}
where $N_{d}=\sum_{k=1}^{K}\mathds{1}(\mathcal{J}_{q(k)})$ denotes the number of targets successfully detected.
For each SNR level, we calculate the average angular error $E$ over all test vectors with at least one successful detection, and then take the square root of this average to obtain the RMSE. In our experiments, we set the parameters $(\delta_{1}, \delta_{2})=(2, 0.6)$. In addition, we report the performance with respect to the NMSE.
\section{Results}
\label{results}
\subsection{Training Losses}
\begin{figure*}
    \centering
    % Row 1, Left: (a)
    \begin{minipage}[t]{0.4\textwidth}
        \centering
        \includegraphics[width=\linewidth]{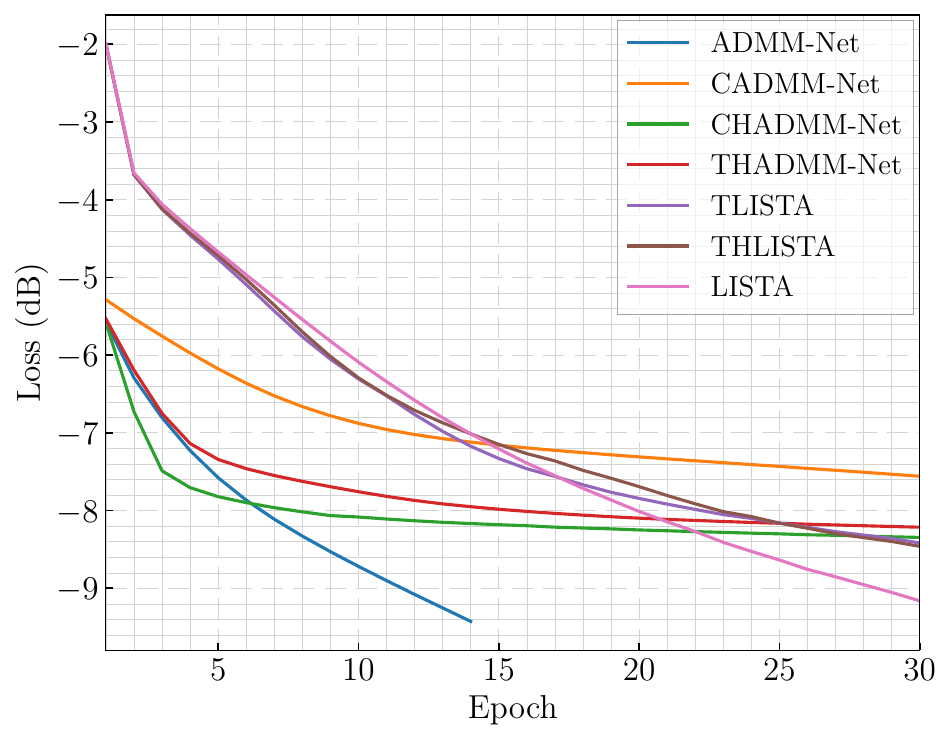}
        \centerline{\small (a)}
    \end{minipage}
    \hspace{0.04\textwidth}
    % Row 1, Right: (b)
    \begin{minipage}[t]{0.4\textwidth}
        \centering
        \includegraphics[width=\linewidth]{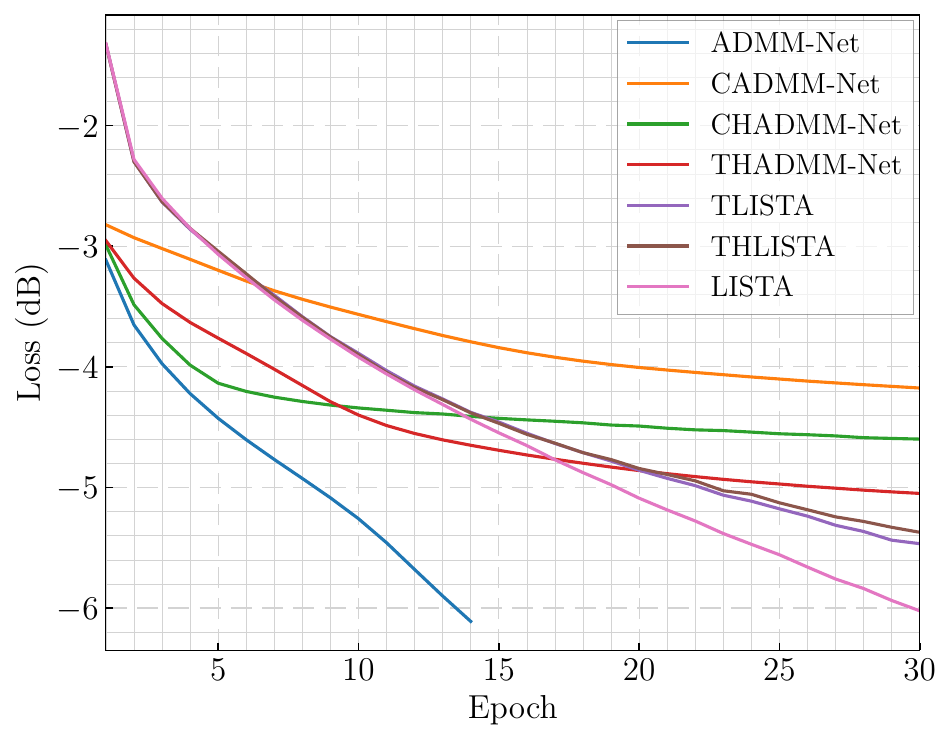}
        \centerline{\small (b)}
    \end{minipage}
    
    \vspace{0.4cm}
    
    % Row 2, Left: (c)
    \begin{minipage}[t]{0.4\textwidth}
        \centering
        \includegraphics[width=\linewidth]{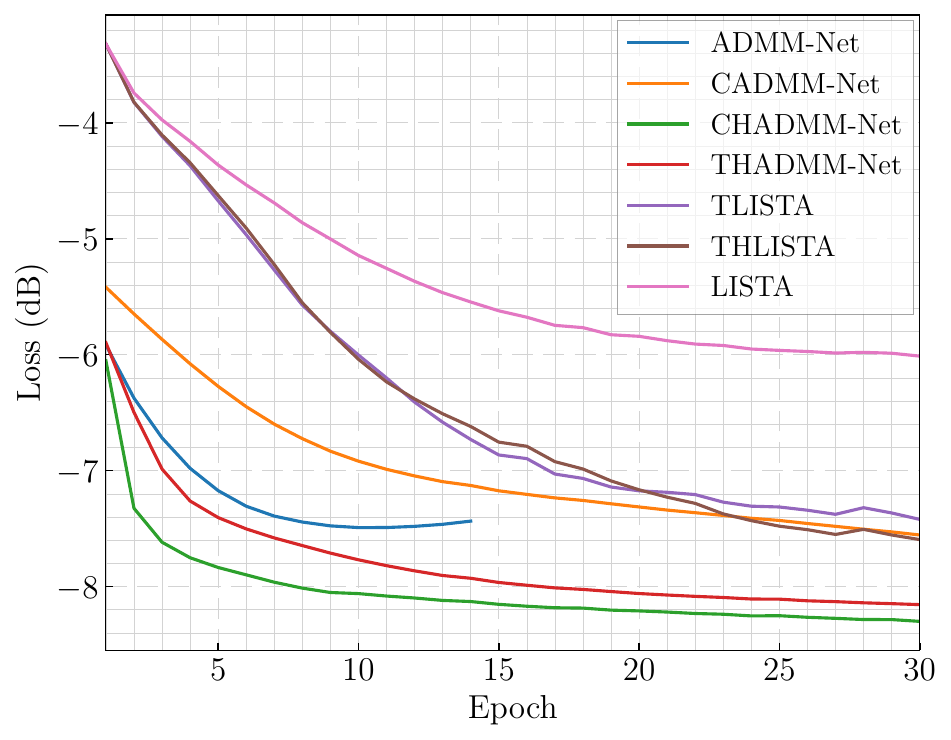}
        \centerline{\small (c)}
    \end{minipage}
    \hspace{0.04\textwidth}
    % Row 2, Right: (d)
    \begin{minipage}[t]{0.4\textwidth}
        \centering
        \includegraphics[width=\linewidth]{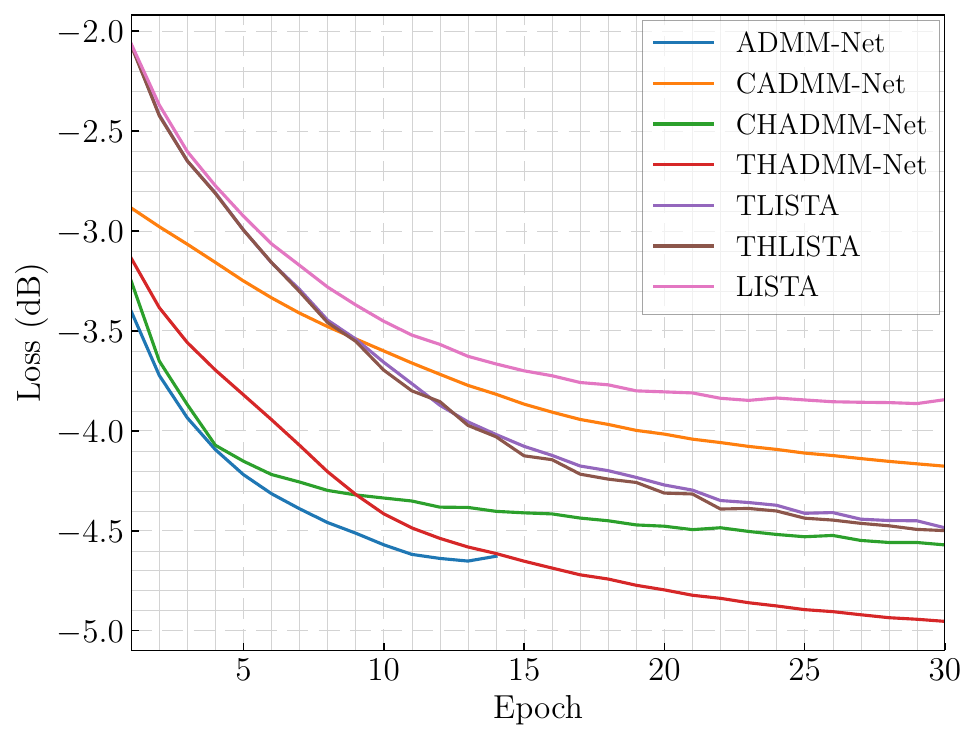}
        \centerline{\small (d)}
    \end{minipage}
    
    \vspace{0.3cm}
    \caption{Training and validation losses for different settings: \textbf{a} training loss (SLA), \textbf{b} training loss (ULA), \textbf{c} validation loss (SLA), and \textbf{d} validation loss (ULA).}
    \label{fig:losses}
\end{figure*}

The training curves in Fig. \ref{fig:losses} indicate that the structured versions tend to learn faster than the unstructured ones. Notably, we can see from Fig. \ref{fig:losses}a that CHADMM-Net under the SLA setting has most of its validation loss decrease during the first 10 epochs only. This stands in sharp contrast with the learning behavior of ADMM-Net, which begins to overfit after the 13-th epoch and thus requires early-stopping \cite{earlystop} to avoid a degradation in performance. One additional observation worth noting is that, despite LISTA having more parameters than ADMM-Net, it does not exhibit the same overfitting pattern, which indicates that the learning behavior of deep-unfolded networks can be largely influenced by the choice of the iterative method being unfolded, as LISTA and ADMM-Net are architecturally different. 
\par Finally, we can note that the networks are able to learn better under the SLA setting compared to the ULA one, which can in part be explained with the fact that the SLA setting, with an identical number of elements but larger aperture, is more favorable for the recovery of sparse signals with closely spaced frequencies.

\subsection{Performance Comparison}

\begin{sidewaysfigure*}
    \centering
    % ================= ROW 1: SLA Scenario =================
    \begin{minipage}[t]{0.31\linewidth}
        \centering
        \includegraphics[width=\linewidth]{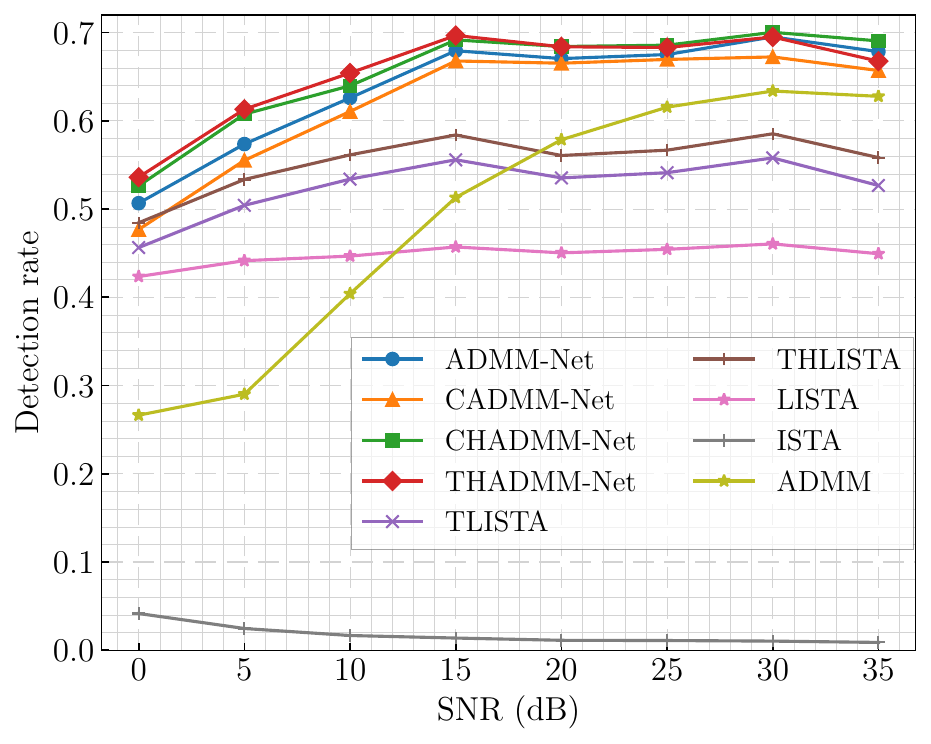}
        \centerline{\small (a)}
    \end{minipage}
    \hfill
    \begin{minipage}[t]{0.31\linewidth}
        \centering
        \includegraphics[width=\linewidth]{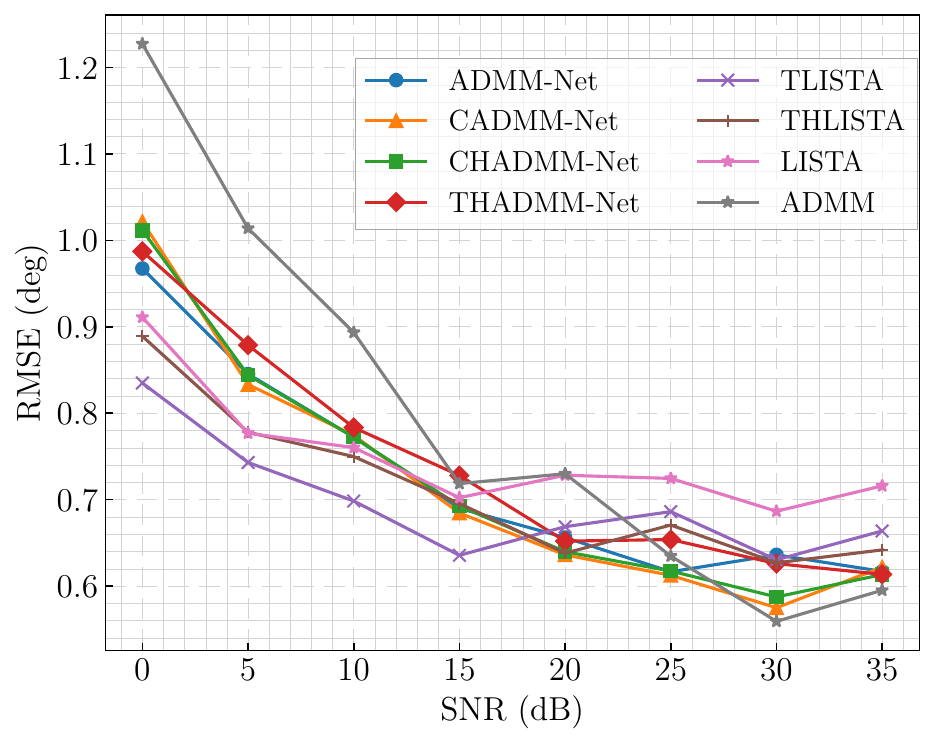}
        \centerline{\small (b)}
    \end{minipage}
    \hfill
    \begin{minipage}[t]{0.31\linewidth}
        \centering
        \includegraphics[width=\linewidth]{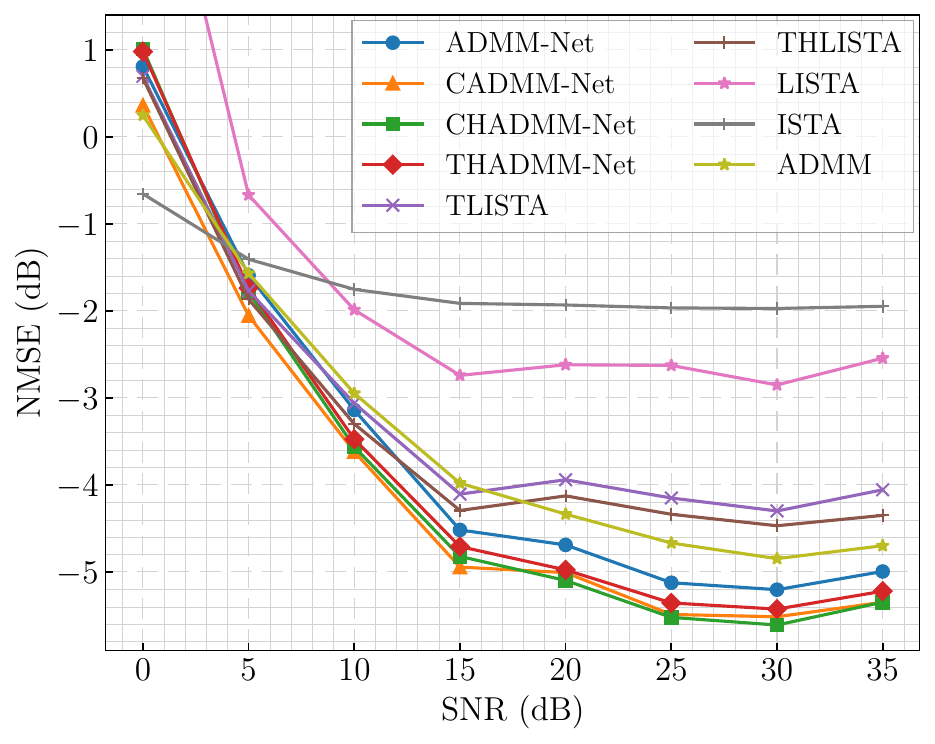}
        \centerline{\small (c)}
    \end{minipage}
    
    \vspace{0.6cm} % Slightly more space since landscape has extra height
    
    % ================= ROW 2: ULA Scenario =================
    \begin{minipage}[t]{0.31\linewidth}
        \centering
        \includegraphics[width=\linewidth]{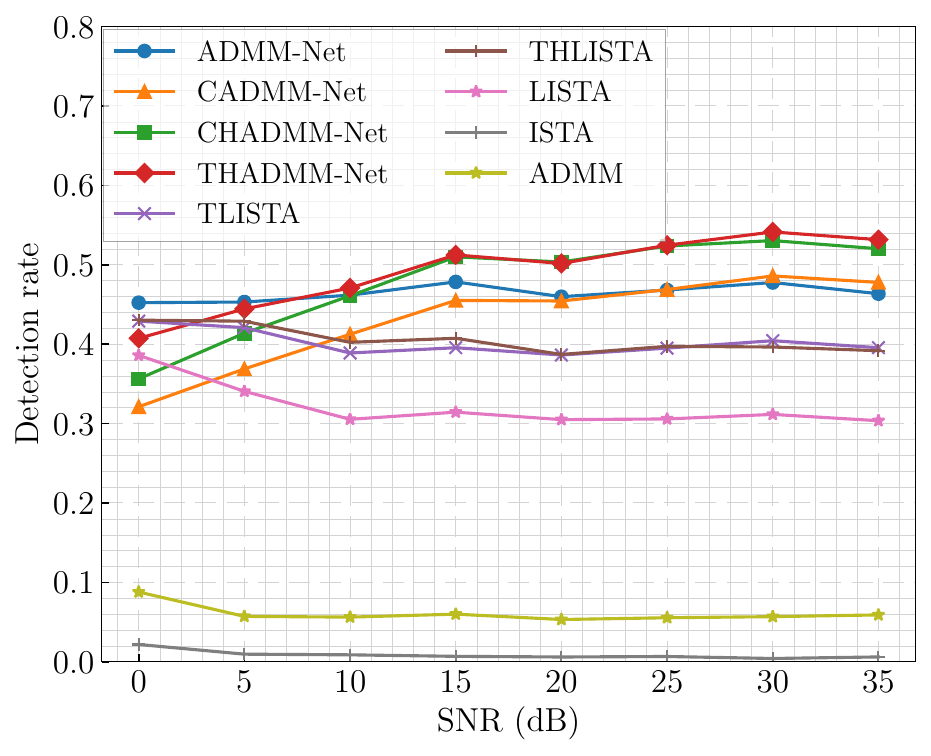}
        \centerline{\small (d)}
    \end{minipage}
    \hfill
    \begin{minipage}[t]{0.31\linewidth}
        \centering
        \includegraphics[width=\linewidth]{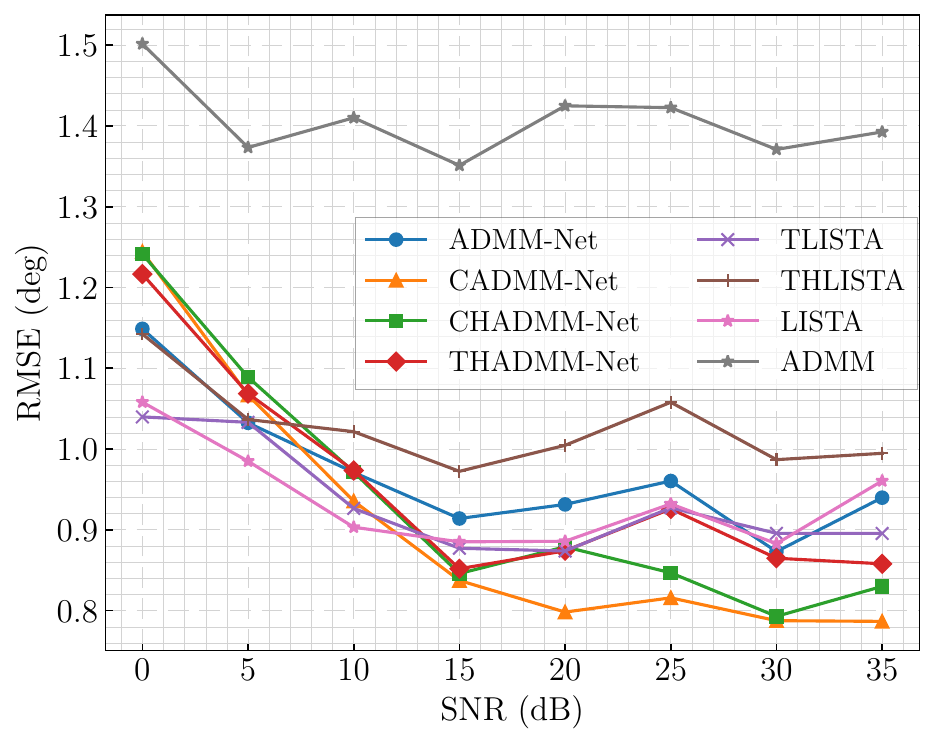}
        \centerline{\small (e)}
    \end{minipage}
    \hfill
    \begin{minipage}[t]{0.31\linewidth}
        \centering
        \includegraphics[width=\linewidth]{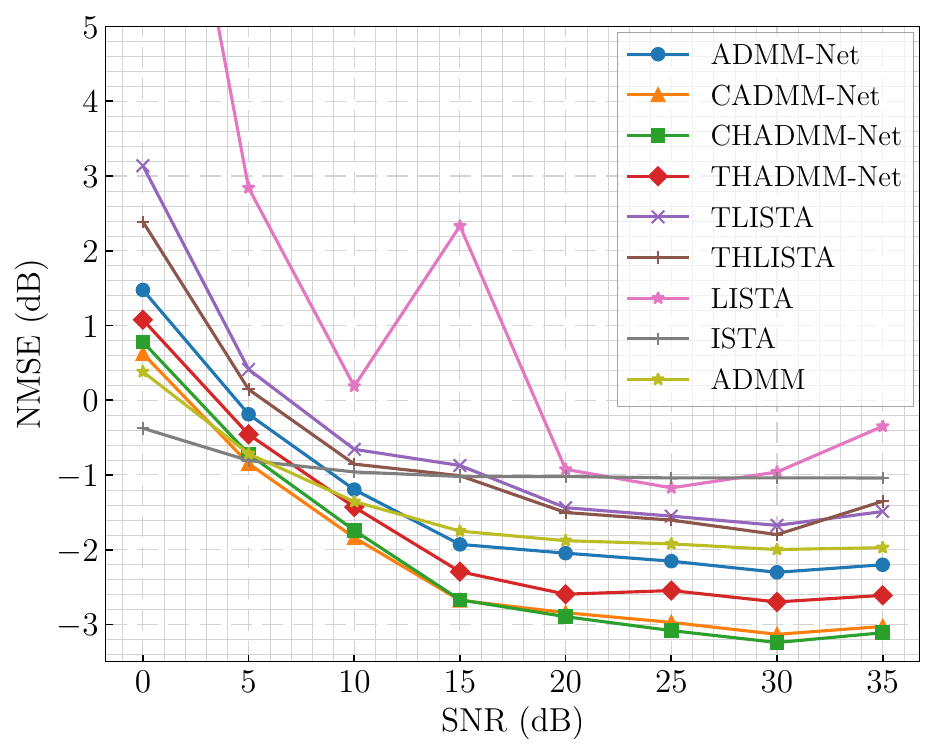}
        \centerline{\small (f)}
    \end{minipage}
    
    \vspace{0.4cm}
    \caption{Performance comparison for the 30-element SLA (top row) and ULA (bottom row) scenarios: \textbf{a} SLA detection rate, \textbf{b} SLA RMSE, \textbf{c} SLA NMSE, \textbf{d} ULA detection rate, \textbf{e} ULA RMSE, and \textbf{f} ULA NMSE. All networks are 30-layers deep. }
    \label{fig:metrics}
\end{sidewaysfigure*}

Fig.~\ref{fig:metrics}a and Fig.~\ref{fig:metrics}d show the detection rate performance of the networks along with a 30 iteration of ADMM and ISTA. We can note that CHADMM-Net and THADMM-Net maintain the lead in the SLA setting across the full SNR range, whereas in the ULA setting we can see that CADMM-Net outperforms the other networks for under 10 dB of SNR, whereas CHADMM-Net and THADMM-Net dominate above that level. We can also note the poor performance of ISTA and ADMM, whereby we can see that for the same number of iterations as the network depth, the deep-unfolded versions outperform their classical iterative counterparts by a significant margin. For the RMSE performance, we can note from Fig.~\ref{fig:metrics}b and Fig.~\ref{fig:metrics}e that all the structured versions exhibit a roughly comparable performance, which is to be expected given that the RMSE is evaluated only after the detection phase. For the NMSE performance, we can note  from Fig.~\ref{fig:metrics}c and Fig.~\ref{fig:metrics}f  that both CHADMM-Net and CADMM-Net exhibit a very close performance to that of ADMM-Net for the SLA setting throughout the SNR range, whereas the performance gap widens in favor of CHADMM-Net to roughly 1 dB for SNRs above 15 dB in the ULA case. 

\par
\begin{figure*}
    \centering
    % ================= ROW 1: SLA Scenario =================
    \begin{minipage}[t]{0.45\textwidth}
        \centering
        \includegraphics[width=\linewidth]{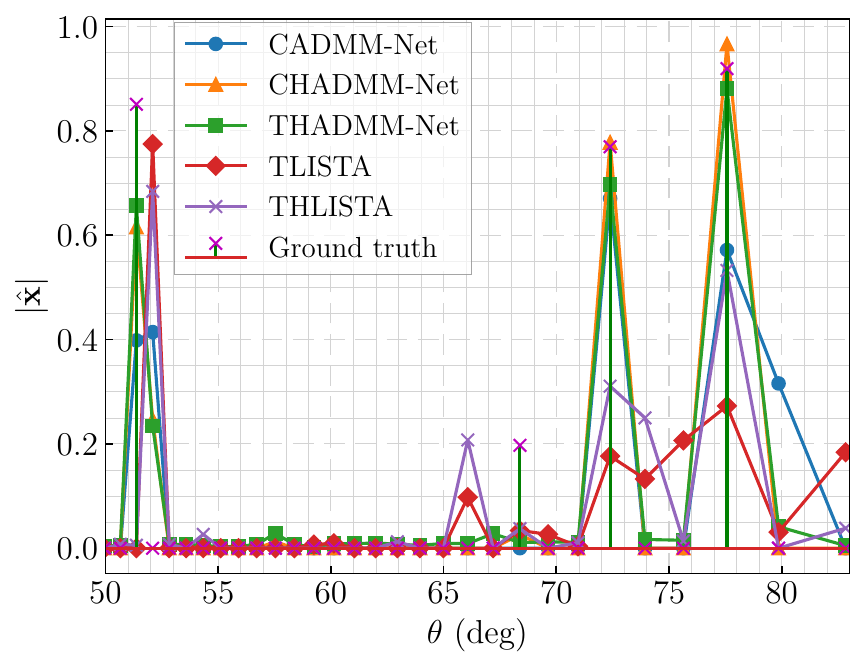}
        \centerline{\small (a)}
    \end{minipage}
    \hspace{0.04\textwidth}
    \begin{minipage}[t]{0.45\textwidth}
        \centering
        \includegraphics[width=\linewidth]{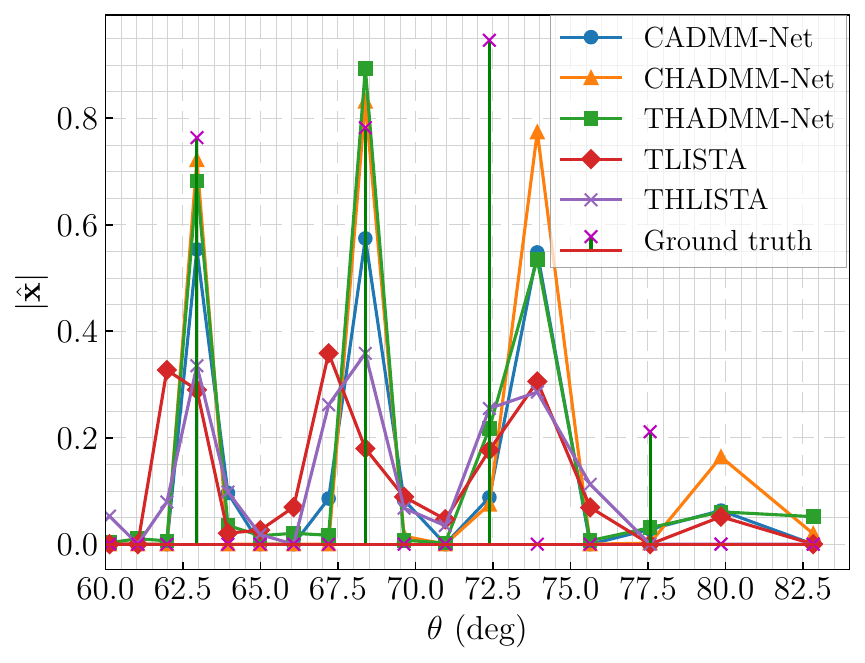}
        \centerline{\small (b)}
    \end{minipage}
    
    \vspace{0.5cm}
    
    % ================= ROW 2: ULA Scenario =================
    \begin{minipage}[t]{0.45\textwidth}
        \centering
        \includegraphics[width=\linewidth]{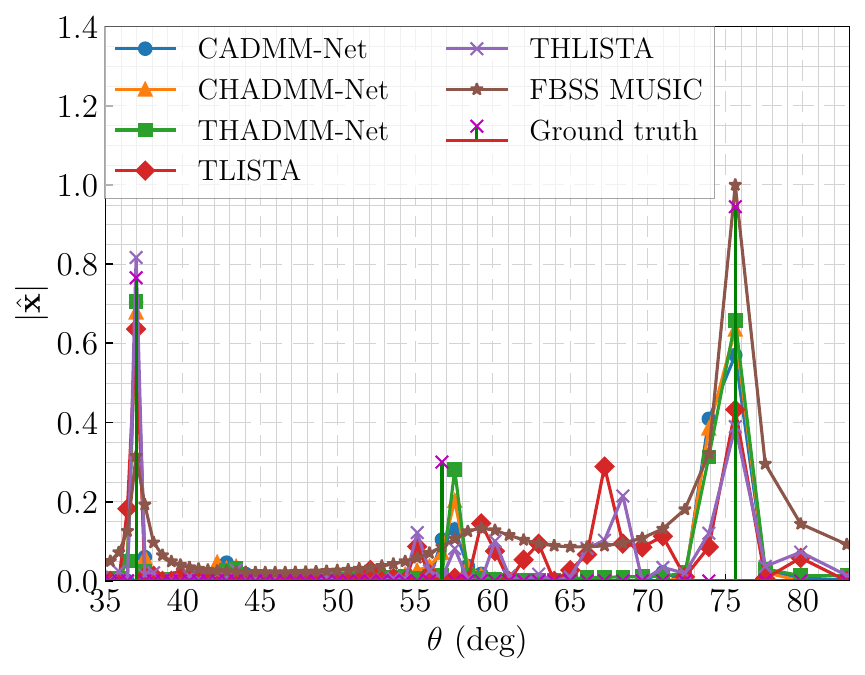}
        \centerline{\small (c)}
    \end{minipage}
    \hspace{0.04\textwidth}
    \begin{minipage}[t]{0.45\textwidth}
        \centering
        \includegraphics[width=\linewidth]{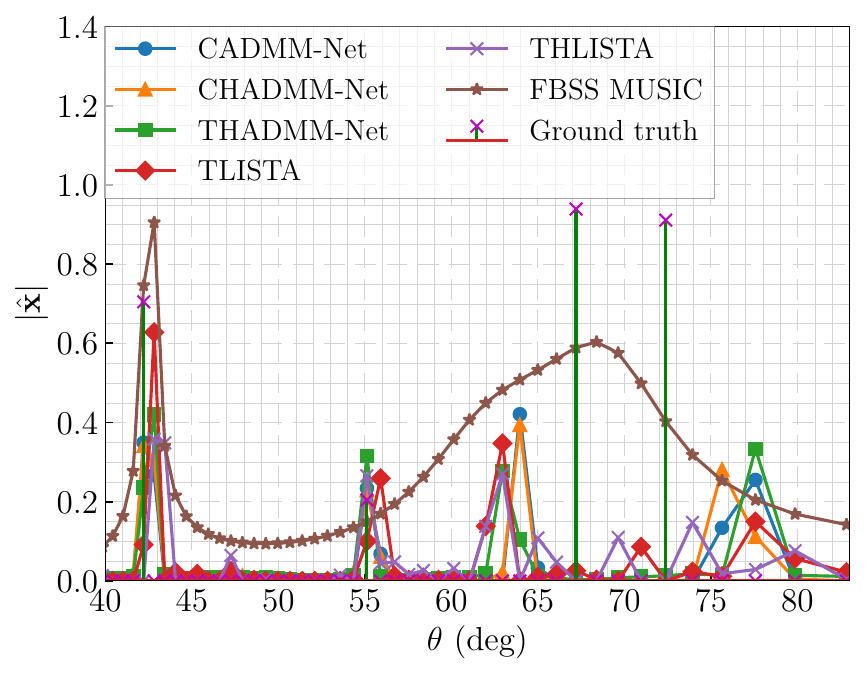}
        \centerline{\small (d)}
    \end{minipage}
    
    \vspace{0.4cm}
    \caption{Sample spectrums at 10 dB SNR: \textbf{a}, \textbf{b} SLA scenario with 60 $\mathbf{\lambda}/\mathbf{2}$ aperture and 30 elements; \textbf{c}, \textbf{d} ULA scenario with 30 elements. The subarray size is set to 10 for FBSS MUSIC.}
    \label{fig:spectrums}
\end{figure*}

 Fig. \ref{fig:spectrums}a and \ref{fig:spectrums}b show a sample spectrum for the SLA setting at 10 dB SNR, where we can see that CADMM-Net and THADMM-Net are able to match to a closer extent both the amplitude and location of the ground truth targets compared to TLISTA and THLISTA. The same can be observed in the ULA setting from Fig. \ref{fig:spectrums}c and Fig. \ref{fig:spectrums}d in which we added a forward-backward spatially smoothed MUSIC spectrum for reference and where the true number of targets is directly provided. We can see that CHADMM-Net and CADMM-Net are able to match THADMM-Net and outperform TLISTA, THLISTA and FBSS MUSIC in terms of detectability and localization error.

\section{Conclusion}
\label{conclusion}
We presented CADMM-Net and CHADMM-Net, two deep-unfolded versions of the ADMM algorithm for DoA estimation, wherein we impose circulant and Hermitian-circulant constraints on the learnable matrices, motivated by the circulant nature of the Gram matrix that appears in the ADMM algorithm under mild assumptions with respect to the array geometry and the uniform nature of the normalized frequency grid used in the dictionary. By exploiting the well-known properties of circulant matrices, we then show that we can considerably reduce the number of learnable parameters per layer as well as decrease the computational complexity. We then performed an extensive characterization of the proposed architectures with respect to the detection rate, the RMSE, and the NMSE, and showed that, in addition to providing both a smaller inference CFLOPS count through the use of FFTs as well as a significantly smaller model parameter count, they are able to maintain a competitive performance against LISTA, ADMM-Net, THADMM-Net, TLISTA, and THLISTA  as well as ISTA and ADMM. 

\newpage
%% BioMed_Central_Bib_Style_v1.01

% \input{sn-article.bbl}

\end{document}